# A two-stage model for dealing with temporal degradation of credit scoring

Maria Rocha Sousa[1], João Gama[1, 2], Manuel J. Silva Gonçalves
[1]Faculty of Economics
[2]LIAAD-INESC Porto
University of Porto
Porto, Portugal

*Abstract* - This work is attached to the BRICS 2013 competition. We propose a two-stage model for dealing with the temporal degradation of credit scoring models. This methodology produced motivating results in a 1-year horizon. We anticipate that it can be extended to other applications of risk assessment with great success. Future extensions should cover predictions in larger time frames and consider lagged periods. This methodology can be further improved if more information about the economic cycles is integrated in the forecasting of default.

## I. INTRODUCTION

In retail banking, credit risk assessment often relies in credit scoring models developed with supervised learning methods used to evaluate a person's credit worthiness, so-called scoring or PD models[1]. The output of these models is a score that translates a probability of a customer becoming a defaulter, usually in a fixed future period. Nowadays, these models are at the core of the banking business, because they are imperative in credit decision-making, in price settlement, and to determine the cost of capital. Moreover, central banks and international regulation have dramatically evolved to a structure where the use of these models is implicit, to achieve soundness standards for credit risk valuation in the banking system.

Since 2004, with the worldwide implementation of regulations issued by the Basel Committee on Banking Supervision within Basel II, banks were encouraged to strengthen their internal models frameworks for reaching the A-IRB (Advanced Internal Rating Based) accreditation. To achieve this certification, banks had to demonstrate that they were capable of accurately evaluating their risks, complying with Basel II requirements, by using their internal risk models' systems, and keep their soundness. Banks owning A-IRB accreditation gained an advantage over the others, because they were allowed to use lower coefficients to weight the exposure of credit at risk, and benefit from lower capital requirements. A lot of improvements have been made in the existing rating frameworks, extending the use of data mining tools and artificial intelligence. Yet, this may have been bounded by a certain unwillingness to accept less intuitive algorithms or models going beyond standard solutions being implemented in the banking industry, settled in-house or delivered through analytics providers (e.g. FICO, Experian, PwC and KPMG).

Developing and implementing a credit scoring model can be time and resources consuming – easily ranging from 9 to 18 months, from data extraction until deployment. Hence, it is not rare that banks use an unchanged credit scoring model for several years. Bearing in mind that models are built using a sample file frequently comprising 2 or more years of historical data, in the best case scenario, data used in the models are shifted 3 years away from the point they will be used. However, to our knowledge an 8-year shift is frequently exceeded. Should conditions remain unchanged, then this would not significantly affect the accuracy of the model, otherwise, its performance can greatly deteriorate over time. The recent financial crisis came to confirm that financial environment greatly fluctuates, in an unexpected manner, posing renewed attention regarding scorecards built upon frames that are by far outdated. By 2007-2008, many financial institutions were using stale scorecards built with historical data of the early-decade. The degradation of stationary credit scoring models is an issue with empirical evidence in the literature [1, 2], however research is still lacking more realistic solutions.

Dominant approaches usually stand on static learning models. However, as the economic conditions evolve in the economic cycle, either deteriorating or improving, also varies the behavior of an individual, and his ability of repaying his debt. Hence, default needs to be regarded as time changing. Also the default evolution echoes trends of the business cycle, and related with this, regulatory movements, and interest rates fluctuations. In good times, banks and borrowers tend to be overoptimistic about the future, whilst in times of recession banks are swamped with defaulted loans, high provisions, and tighten capital buffers turn highly conservative. The former leads to more liberal credit policies and lower credit standards, the later promotes sudden credit-cuts. Empirical evidence and theoretical frameworks support a positive, and lagged relationship between rapid credit growth and loan losses. Within this context, public studies are mostly motivated by credit consumption and regulatory concerns, and therefore concentrated to model macroeconomic factors. So far, none has explicitly integrated them with factors of specific risk in the existing credit scoring for retail finance.

---
[1] Other names can be used to refer to PD models, namely: credit scoring, credit risk models, scorecards, credit scorecards, rating systems or rating models, although some have different meanings.

Traditional systems that are one-shot, fixed memory-based, trained form fixed training sets, and static models are not prepared to process the highly detailed evolving data. And so, they are not able to continuously maintain an output model consistent with the actual state of nature, or to quickly react to changes [3]. These are some of the features of classic approaches that put evidence on that the existing credit scoring systems are limited. As the processes underlying credit risk are not strictly stationary, consumers' behavior and default can change over time in unpredictable ways. There are several types of evolution inside a population, like population drifts, that translate into changes in the distributions of the variables, affecting the performance of the models. There is a new emphasis on running predictive models with the ability of sensing themselves and learn adaptively [3]. Advances on the concepts for knowledge discovery from data streams suggest alternative perspectives to identify, understand and efficiently manage dynamics of behavior in consumer credit in changing ubiquitous environments. In a world in which events are not preordained and little is certain, what we do in the present affects how events unfold in unexpected ways.

This paper follows in section II with a brief description of the problem; the objectives and the database are succinctly presented. Section III details the methodology and theoretical framework of this research. A one-dimensional analysis is presented, as well as an overall assessment of the data available for modeling. A multidimensional approach is exposed in section IV where we propose a two-stage model for dealing with the temporal degradation of credit scoring. In the first stage we develop a credit scoring, by comparing several supervised learning methods. In the second stage, we introduce the effect of time changing environment by shifting the initial predictions according to a factor of the expected variation of default. Results in a 1-year horizon are presented in section V. Conclusions and future applications of the two-stage model are discussed in Section VI.

## II. THE PROBLEM

### A. Objectives

This research aims to propose a new approach for dealing with the temporal degradation of a portfolio of customers with credit cards in a financial institution operating in Brazil. Our work is attached to the BRICS 2013 competition, and is based in a real world data set, along two years of operation, from 2009 to 2010. This competition was riven in two tasks[2], each of them focused on two features of the credit risk assessment model:

- Task 1: Develop a scorecard, tilting between the robustness in a static modeling sample and the performance degradation over time, potentially caused by market gradual changes along few years of business operation.

- Task 2: Fitting of the estimated delinquency produced by an estimation model to that observed on the actual data for the applications approved by the scorecard.

Participants were encouraged to use any modeling technique, under a temporal degradation or concept drift perspective. The official performance metrics were the area under the ROC curve for task 1, and the Chi-square for the monthly estimates of delinquency in task 2, provided that the average delinquency is kept within half and double of the actual delinquency in the period. Innovative ways of handling task 1 can be found in PAKDD 2009 Competition whose focus was on this type of degradation. Task 2 represents an innovation in data mining competitions worldwide by emphasizing the relevance of the quality of future delinquency estimation instead of the usual lowest future average delinquency. Our approach was to build an integrated solution to deal with these two tasks. First we developed a credit scoring using a set of supervised learning methods. Then we calibrated the output, based on a projection of the evolution in the default. This forecast considered both the evolution of the default and the evolution of exogenous data series, echoing potential changes in the population of the model, in the economy, or in the market. In so doing, resulting adjusted scores translate a combination of the customers' specific risk with systemic risk. Theoretical models for dealing with knowledge discovery from data streams sound suitable for dealing with these problems. However, as in many other applied financial studies, this research is bounded by some practical limitations, like systematic noise in the data and short time series. Our view is that the choice of the most appropriate methods for developing a credit scoring is context specific. Therefore, we present a technical approach mostly driven by the specifics of the problem. We also take in consideration the extent of meaningful and reliable data that is actually available for modeling.

### B. Database and competition

The research summarized here was conducted in a real-life financial dataset, comprising 762,966 records, from a financial institution in Brazil. Data for modeling was provided along two years of operation, from 2009 to 2010. Each customer in the modeling dataset is assigned to a delinquency outcome - good or bad. In this problem, a person is assigned to the bad class if she had a payment in delay for 60 or more days, along the first year after the credit has been granted. The delinquency rate in the modelling dataset is 27.3%. Two additional datasets from the subsequent year, 2011, were used to test the performance achieved in the static modelling sample – the leaderboard and the prediction dataset. The leaderboard contains a sample of 60,000 records that were obtained aggregating subsamples of 5,000 applications of each month in 2011. Although the default outcome was not available in the leaderboard dataset, the discriminatory power of the model given by the receiver operating characteristic curve (AUC) could be known during the modeling stage. Competitors were allowed to make submissions of their solutions in the leaderboard, and for each the area under the ROC curve and the distance D were delivered online. The prediction dataset will be used for the final performance evaluation in the competition. This dataset has 444,828 applications in 2011, for which the default outcome was not available at any stage of the modeling. A summary of the files is presented in TABLE I.

---
[2] We made some modifications to the original text describing the two tasks, in order to reflect our understanding of the aim of the competition, which is at the basis of the proposed solution.

TABLE I. DATASETS SUMMARY

| Dataset [a] | Records | Period | Target | Delinquency (%) |
|---|---|---|---|---|
| Modeling | 762,966 | 2009-2010 | Labeled | 0.273 |
| Leaderbord | 60,000 | 2011 | Unlabeled | --- |
| Prediction | 444,828 | 2011 | Unlabeled | --- |

The full list of variables in the original data set is available in the BRICS 2013 official website. It contains 39 variables, categorized in TABLE II and one target variable with values 1 identifying a record in the bad class and 0 for the good class.

TABLE II. PREDICTIVE VARIABLES SUMMARY

| Type | # | Information |
|---|---|---|
| Numerical | 6 | Age, monthly income, time at current address, time at current employer, number of dependents, and number of accounts in the bank. |
| Treated as nominal | 13 | Credit card bills due date, 1st to 4th zip digit codes, home (state, city, and neighborhood), marital status, income proof type, long distance dialing code, occupation code, and type of home. |
| Binary | 16 | Address type proof, information of the mother and father's names, input from credit bureau, phone number, bills at the home address, previous credit experience, other credit cards, tax payer and national id, messaging phone number, immediate purchase, overdraft protection agreement, lives and work in the same state, lives and work in the same city, and gender. |
| Date | 1 | Application date. |
| ID | 3 | Customer, personal reference, and branch unique identifiers. |

### III. METHODOLOGY AND THEORETICAL FRAMEWORK

This research evolves from a one-dimensional analysis, where we come across the financial outlook underlying the problem, to a multidimensional approach, where we gradually develop and experiment a new framework to model credit risk. The one-dimensional analysis was tailored to gain intuition on the default predictors and the main factors ruling the dynamics of default. The multidimensional approach is at the core of our work and was held in two stages. In the first stage we shaped a credit scoring model from a classical framework, with a static learning setting and binary output. In the second stage, we used a linear regression between exogenous data and the internal default for adjusting the predictions of default given by the credit scoring model.

Exogenous data series from the Central Bank of Brazil [4] and the Brazilian Institute of Geography and Statistics [5] were also considered for evaluating the effect of time changing economics in the default evolution. We used quarterly series from January 2004 to December 2011. Nevertheless, for determining the fitting between them and the internal default, only the period 2009 to 2010 was considered. We used the coefficient of determination, r-square, to evaluate the fitting between these series and the internal default, in a one-dimensional basis.

#### A. Task 1 – Data analysis, cleansing and new characteristics

Some important aspects of the datasets were considered, because they can influence the performance in the unlabeled datasets. These aspects regard to:

*Great extent of zero or missing values* – In exception to the variables 'lives and work in the same state' and 'previous credit experience', binary flags have 95% to 100% concentrated in one of the values, which turn them practically unworkable. The same occurs for the numerical variables 'number of dependents' and 'number of accounts in the bank', both with more than 99% zeroes. The remaining variables were reasonably or completely populated.

*Outliers and unreasonable values* – The variable age present 0.05% of applications assigned to customers with ages between 100 and 988 years. An immaterial fraction of values out of the standard ranges are observable in the variables credit card bills due day, monthly income and time at current employer.

*Unreliable and informal information* – Little reliability on socio-demographic data is amplified by specific conditions in the backdrop of this problem. This type of scorecards is usually based in verbal information that the customer provides, and in most of the cases no certification is made available. In 85% of the applications, no certification for the income was provided, and 75% do not have proof for the address type. Customers have little or no concern to provide accurate information. The financial industry is aware of this kind of limitations. However, in highly competitive environments there is little chance to amend them, while keeping in the business. Hence, other than regulatory imperatives, no player is able to efficiently overcome data limitations. As currently there are no such imperatives in Brazilian financial market, databases attached to this type of models are likely to keep lacking reliability in the near future.

*Bias on the distributions of modeling examples* – The most noticeable bias is in the variable monthly income. Values shift from one year to another. This is most likely related to increases in the minimum wages and inflation. Slight variations are also observable in the geographical variables, which are possibly related with the geographical expansion of the institution. In the remaining characteristics, the correlation between the frequency distributions of 2009 and 2010 range from 99% to 100%, suggesting a very stable pattern during the analyzed period.

*Data cleansing and new characteristics* - We focused the data treatment on the characteristics that were reasonably or fully populated. Fields state, city, and neighborhood contain free text, and were subjected to a manual cleansing. Classes with 100 or less records were assigned to a new class "Other". We could observe that there may be neighborhoods with the same name in different cities; and hence we concatenated these new cleansed fields into a new characteristic. Taking into account that the shift in the variable monthly is likely related to an increase of the minimum wages and inflation, a new characteristic was calculated by multiplying the inflation rate in the year by the variable monthly income.

*Data transformation* - Variables were transformed using the weights of evidence (WoE) in the complete modeling dataset. $WoE = ln\left(\frac{g/G}{b/B}\right)$, where g and b are respectively the number of good and the number of bad in the attribute, and G and B are respectively the total number of good and bad in the population sample. The larger the WoE the higher is the proportion of good customers. For the nominal and binary variables we calculated the WoE for each class. Numerical variables were firstly binned using SAS Enterprise Miner, and then manually adjusted to reflect the domain knowledge. In so doing we aim to achieve a set of characteristics less exposed to overfitting. Cases where the calculation of the WoE rendered impossible - one of the classes without examples - were given an average value. The same principle was applied to values out of the expected ranges (e.g. credit card bills due day higher than 31).

*One-dimensional analysis* - The strength of each potential characteristic was measured using the information value (IV)

$$IV = \sum_{i=1}^{n}\left(g/G - b/B\right)WoE,$$ where n is the number of attributes in the characteristic. The higher is the IV the higher is the relative importance of the characteristic. In a one-dimensional basis, the most important characteristics are age, occupation, time at current employer, monthly income and marital status, with information values of 0.368, 0.352, 0.132, 0.117, and 0.116, respectively. Remaining characteristics have 0.084 or less.

*Interaction terms* - Using the odds in each attribute of the variables, we calculated new nonlinear characteristics using interaction terms between variables to model the joint effects. We tested six combinations, for which we present the IV in TABLE III.

TABLE III. INFORMATION VALUES FOR THE TESTED COMBINATIONS

| Combination | IV |
|---|---|
| Age ∗ Income | 0.315 |
| Age ∗ Occupation | 0.009 |
| Income ∗ Marital status | 0.208 |
| Income ∗ Occupation | 0.334 |
| Income ∗ Proof of income | 0.123 |
| Age ∗ Income ∗ Occupation | 0.007 |

### B. Task 2 - changing environment and time series analysis

This work aims to propose an innovation in the ground of data mining for credit scoring, by fitting the delinquency estimated with a scorecard, based on the motion of specific factors in the financial and economic environments. We based our empirical study in the analysis of exogenous time series, TABLE II. Among the exogenous series that are available, we expect that this set may exhibit a major influence in the behaviour of private individuals and in their pattern of default with credit cards.

TABLE IV. EXOGENOUS DATA SERIES FOR BRAZIL

| Series | Correl[a] | R square |
|---|---|---|
| Default/financial revenue in credit cards in Brazil[b] | 0.805 | 0.648 |
| Default in revolving credit in Brazil[b] | 0.491 | 0.241 |
| GDP | -0.168 | 0.028 |
| GDP annual variation | 0.485 | 0.236 |
| Primary income payments (BoP, cureent US$) | 0.787 | 0.619 |
| Lending interest rate (%) | 0.118 | 0.014 |
| Real interest rate (%) | -0.047 | 0.002 |
| Taxes on income, profits and capital gains (% of revenue) | -0.256 | 0.065 |
| Household final consumption expenditure (% of GDP) | -0.713 | 0.509 |
| Private consumption | -0.365 | 0.133 |
| Inflation rate | 0.797 | 0.635 |
| Unemployment rate | -0.100 | 0.010 |
| Consumer confidence | -0.280 | 0.080 |
| Wages | 0.381 | 0.145 |

[a.] Correlation between the internal default and the exogenous data series, from 2009 to 2010.
[b.] An abnormal observation was recorded in the first quarter of 2011. As we could not confirm the reliability of this value, we opted to replace it by the average of the contiguous quarters.

As most of the available exogenous data series are quarterly updated, we considered quarterly points for all the series. Although exogenous data series are available from 2004 onwards, we could not make a full use of them, because the internal data was available just for 2009 and 2010. Hence, we focused the analysis on that period, which we consider short for achieving a reliable forecast. A minimum of 5 years is required with Basel II. In this type of analysis, it would be appropriate using a much larger historical period, to capture different phases of one or more economic cycles[3]. It follows that two years are clearly scarce. However, as this competition aimed for an accurate 1-year forecast, our assumption was to consider that in 2011, Brazil would be in the same phase of the economic cycle as in 2009 and 2010. The internal default series follow very different paths in 2009 and 2010, Fig. 1, which discourages any attempt to discern an intra-annual seasonality. Nevertheless, in both years, the default slightly increased along the second semester. Although this is not firmly conclusive, we considered this occurrence in the forecasting scenarios, as we will describe hereafter.

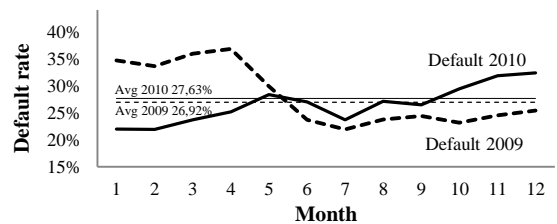

Fig. 1. Internal default by month along 2009 and 2010.

---

[3] An economic cycle may last for decades. Identifying an economic cycle is an important and non-trivial task that will be no further analysed here, as it goes beyond the scope of this work.

In order to find potential relations between the internal default and the exogenous series, we calculated their correlations, TABLE IV. For the analysed period, the best series are the default on the financial revenues of credit cards in Brazil, primary income payments, households' final consumption expenditure, and the inflation rate. As only 8 observations are available, linear regression should consider a single independent variable, aiming to avoid overfitting. In forecasting scenarios, we only considered the series with the the highest correlation - default on the financial revenues of credit cards in Brazil. Notwithstanding, an r-square of 64% in the regression can be considered low for the regression. We tested three forecasting scenarios, summarized in TABLE V, which were iteratively submitted to the leaderboard. The final prediction is based on the scenario with the lowest distance D in the leaderboard.

TABLE V. FORECAST SCENARIOS TESTED IN THE TASK 2

| # | Scenarios Description | Rational |
|---|---|---|
| 1 | Estimate the default in each quarter of 2011 adaptively from the values of default in credit cards of the previous quarter, and submit calculated values. | Incorporate new information adaptively, when it is available. This may benefit from the drift detection and suggest implementing corrective actions. |
| 2 | Estimate the default in each quarter of 2011 adaptively from the values of default in credit cards of the previous quarter, and submit the average. | As there is no knowledge about the economic cycle, any correction resulting from a drift between quarters are unsubstantiated in this application. It is more appropriate using a central tendency of default. |
| 3 | Submit the average annual default until September, and the average increased by 1% in the last two months. | Use central tendency of default and adjust the months were the direction of the drift is more certain. |

## IV. TWO-STAGE MODEL

Some previous research suggests including economic conditions directly into a regression scorecard [6], survival analysis [7], or transition models [8]. Our approach was to use a two-stage modeling framework in order to keep apart the two dimensions of risk – specific and systemic. The specific risk should be captured from the bank internal database using a scorecard, and the movement of systemic risk is determined with a linear regression. The final default forecast results from the adjustment of the initial score using a factor of the expected variation in the default of the population of the model.

### A. Task 1: Credit scoring construction

Several standard classification models, based on logistic regression (LR), AdaBoost, and Generalized Additive Models (GAM) were designed and tested with a 10 fold crossed-validation. Characteristics were iteratively added to the models, until there no performance gain was observed in the test. Four different strategies were gauged, varying the window, and eliminating noise from the input set:

*High volume and diversity* - Use the entire modelling dataset (2009-2010), for more volume and diversity.

*Through-the-door* - Use a sample closest to the through-the-door population, for mitigating the effects of the temporal bias. We tested two different windows: 2010 full year, and 2010 last quarter.

*Ensemble of 12 models* - Create a model to apply in each month of the year, using the corresponding months in 2009 and 2011. Our idea is to mitigate the changes in demand that may be attached to seasonal effects. Twelve models were developed, and their results were combined in the final score.

*Cleaning* - To overcome the presence of noise in the data, create a model using the entire modelling dataset (2009-2010); since the presence of noise in the data can spoil the model, we removed the examples strongly misclassified, and retrained the model in the reduced set. We consider an example strongly misclassified if the predicted posterior probability of the true class is less than 0.05.

As the financial institution is expanding, the leaderboard and the prediction sets should contain new codes in the discrete variables (e.g. new branch and geographical codes). These cases cannot be trained from the modelling dataset, because they cannot be observed before the expansion. As there was no knowledge about the nature and strategy of the expansion, we opted to assign an educated guess - use the average partial score for unfamiliar codes. In real-word environments, this assignment is usually better informed, because the strategy is known beforehand.

### B. Task 2: Introduce the effect of time changing environment

In the second stage of modeling we introduced the effect of time changing environment into the scorecard previously developed. The initial predictions of default of the scorecard are shifted according to a factor of the expected variation of default. To any cut-off point set between 0 and 1 corresponds to a new adjusted default rate within the set of approved applications, as illustrated in Fig. 2.

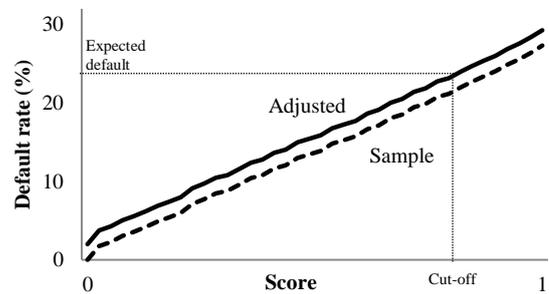

Fig. 2. Scorecard adjustment for the central tendency of default.

## V. RESULTS

*Task 1* - Only the models with the best results in the test set were submitted to the leaderboard. TABLE VI shows the results for the best models in the modelling data set. Although a broad number of different configurations were tested, we only present the results of the models with an AUC higher than 0.71 in the test. This includes the models that were submitted to the leaderboard.

TABLE VI. BEST MODELS IN TASK 1, AUC>0.71 IN THE TEST SET

| Method | Period | Test | Leaderboard[a] |
|---|---|---|---|
| GAM | 2009-10 | 0.7320 | 0.7227 |
| GAM | 2010 | 0.7140 | 0.7131 |
| GAM | 2010(Q4) | 0.7204 | n.a. |
| LR | 2009-10 | 0.7222 | n.a. |
| LR with cleaning | 2009-10 | 0.7222 | n.a. |
| LR monthly | 2009-10, month against month | 0.7230 | 0.7140 |
| AdaBoost | 2009-10 | 0.7180 | n.a. |

[a.] n.a. - not available, because the model was not submitted to the leaderboard.

The model with the best results in the leaderboard is based on the Generalized Additive Model, and using the entire modeling dataset (2009 and 2010). This model was submitted in the final prediction. Although we have tested different timeframes when modeling, it became apparent that for this application, the key for achieving the model with best performance was to bet in higher volume of examples and diversity. The degradation of the proposed scorecard from the test (2009-2010) to the leaderboard (2011) is of 0.93%, which is quite acceptable in real-world applications. The degradation was partly controlled by adjusting unstable variables (e.g. monthly income based on the inflation rate) and by controlling the partial scores for the unfamiliar codes. Apart from these, the bulk of remaining the characteristics is quite stable over time.

*Task 2* – The results of this task, presented in TABLE VII, demonstrate that when the forecast depend on very short time series, the best fitting is achieved with the simplest adjustment – the central tendency of default. In this problem, submitting an average default enhanced the fitting of default (scenario 2), which was further improved by adjusting on the months where the direction of the drift is more certain (scenario 3).

TABLE VII. RESULTS IN TASK 2.

| Scenario (#) | Average default rate | Distance D |
|---|---|---|
| 1 | 0.293 | 3.16 |
| 2 | 0.293 | 1.45 |
| 3 | 0.294 | 0.83 |

[a.] Annual average forecasted default rate in 2011, if the entire portfolio is approved.

## VI. CONCLUSIONS

Theoretical models for knowledge discovery from data streams sound suitable for dealing with temporal degradation of credit scoring models. The idea is to use adaptive models, incorporating new information when it is available. Integrating new information may also benefit from the drift detection, and the occurrence of a drift may suggest eventual corrective actions to the model. Some specifics of the financial problems may turn the models quite stable along time, which is the case of the scorecard presented in this research. A static learning setting was at the basis of the model with the best discriminatory power. It also becomes apparent that some sort of time discretization may turn useless in some applications and may lead to nonsense or suboptimal forecasts. In this problem, as there is no intra-annual seasonality of default, the practical meaning of a monthly prediction is debatable. Credit risk assessment is one area where the data mining and forecasting tools have largely expanded over the last years. However there are a few arenas where the use of these tools should focus essentially on providing a direction, rather than providing a strict prediction. There are a number of possible directions that no model that looks just into the past can enlighten about the future. This includes the directions driven by the business strategy of the bank (e.g. plans for expanding the branch network, to offer a new product, or to merge with another financial institution). The same applies to the occurrence of extreme or rare events, like those that were roused by the recent financial crisis. The new paradigm of forecasting turns out to be looking at hidden streams in the present signal and understand how they will possibly direct an event into the future.

We propose a two-stage model for dealing with temporal degradation of credit scoring, which provided good results in a 1-year timeframe. However, it becomes apparent that this type of models should also prove a good performance in the long run. Therefore, future applications of this modeling framework should be tested in larger timeframes and consider lagged periods. Stress-testing this methodology should consider environments under major macroeconomic distress, or drifting populations resulting from expressive growth of the portfolios of customers.

REFERENCES

[1] J. N. Crook, et al., "The degradation of the scorecard over the business cycle," IMA Journal of Management Mathematics, vol. 4, pp. 111-123, 1992.

[2] A. Lucas, "Updating scorecards: removing the mystique," Readings in Credit Scoring: Foundations, Developments, and Aims. Oxford University Press: New York, pp. 93-109, 2004.

[3] J. Gama, Knowledge discovery from data streams. London: Chapman & Hall/CRC, 2010.

[4] BCB, "Relatório sobre a Indústria de Cartões de Pagamentos Adendo Estatístico," 2011.

[5] www.tradingeconomics.com, 17-06-2013.

[6] M. Zandi, "Incorporating economic information into credit risk underwriting " Credit Risk Modelling Design and Application, pp. 155-158, 1998.

[7] T. Bellotti and J. Crook, "Credit scoring with macroeconomic variables using survival analysis," Journal of the Operational Research Society, vol. 60, pp. 1699-1707, 2008.

[8] M. Malik and L. C. Thomas, "Transition matrix models of consumer credit ratings," International Journal of Forecasting, vol. 28, pp. 261-272, 2012.